# A Decentralized Local Flexibility Market for Local Energy Communities to Mitigate Grid Congestion: A Case Study in Sweden

Maryam Mohiti, Mohammadreza Mazidi, David Steen, Le Anh Tuan, *Member, IEEE*

***Abstract*—This paper proposes a preventive congestion management framework with joint Local Flexibility Capacity Market (LFCM) and Local Energy Markets (LEMs). The framework enables Local Energy Communities (LECs) to optimize their flexibility potential across the LEM, LFCM, and heat markets. The LECs utilize their heat and electricity resources to offer flexibility services to Distribution System Operators (DSOs) for congestion relief. In this framework, energy and flexibility are treated as separate variables, each subject to different pricing scheme. Flexibility prices are market-driven, dynamically reflecting the location and severity of congestion. A case study conducted at Chalmers University of Technology, Sweden, shows that the proposed framework can effectively mitigate congestion by trading the LECs flexibility in the LFCM. The study also highlights up to 40% financial benefits for LECs, promoting the LFCM as a viable solution for congestion management in future decentralized energy systems.**

**Nomenclature**

***Indices***

| | |
|---|---|
| $i, j$ | Index of busses |
| $k$ | Index of iteration |
| $l$ | Index of branches |
| $t$ | Index of time |
| $LEC$ | Set of distribution energy resources (DERs) of LEC |
| $\mathbb{N}^{\boldsymbol{DN}}$ | Set of busses of distribution network (DN) |
| $\mathbb{N}^{\boldsymbol{LEC}}$ | Set of busses of LEC |
| $\mathbb{N}^{\boldsymbol{pcc}}$ | Set of busses that are point of common coupling (PCC) of LECs |

***Variables***

| | |
|---|---|
| $\delta_{t,i}$ | Voltage angle of bus |
| $\lambda_{t,i}^{Flex,k}$ | Flexibility price |
| $\rho_{t,i}^{Congestion,k}$ | Penalty factor of congestion management |
| $E_{t,i}^{DR}/P_{t,i}^{DR}$ | Energy/power of already curtailed with the demand response program |
| $FLEX_{t,i}^{max}$ | Maximum flexibility offered by LECs |
| $LEC_{t,i\in\mathbb{N}^{pcc}}^{Flex}$ | Accepted amount of flexibility from LEC by DSO |
| $H_{t,i}^{B2DH}$ | Heat of boiler fed directly to the heating network |
| $H_{t,i}^{CHP}$ | Output heat of combined heat and power (CHP) unit |
| $H_{t,i}^{ch,TES}/H_{t,i}^{dis,TES}$ | Charge/discharge heat of thermal energy storage (TES) |
| $H_{t,i}^{HP}$ | Output heat of heat pump (HP) |
| $H_{t,i}^{idle,TES}$ | Idle losses of TES |
| $H_t^{Im,LEC}/H_t^{Ex,LEC}$ | Imported/Exported heat to LEC from district heating |
| $H_{t,i}^{SB}$ | Heat input to smart boiler (SB) |
| $H_{t,i}^{R,Stir}$ | Residual heat of Stirling Engine |
| $H_{t,i}^{T}$ | Heat of boiler fed to the CHP turbine |
| $P_{t,der}$ | Active power of DER unit |
| $P_{i,t}^{ch,BES}/P_{i,t}^{dis,BES}$ | Charge/discharge power of Battery energy storage (BES) |
| $P_{t,i}^{CHP}/Q_{t,i}^{CHP}$ | Active/reactive power of CHP unit |
| $P_{t,i}^{DG}/Q_{t,i}^{DG}$ | Active/ reactive power of distributed generation |
| $P_{t,i}^{HP}$ | Electricity input of HP |
| $P_t^{Im,LEC}/P_t^{Ex,LEC}$ | Imported/exported active power to/from LEC |
| $P_{i=0,t}^{grid}/Q_{i=0,t}^{grid}$ | Exchanged active/reactive power with the upstream network |
| $P_{t,i}^{SB}/Q_{t,i}^{SB}$ | Active/reactive power of SB |
| $P_{i,t}^{LS}$ | Involuntary active load shedding |
| $P_t^{Im_base,LEC}/P_t^{Ex_base,LEC}$ | Import/export power of LEC to DN in baseline schedule |
| $P_{t,i,j}^{flow}/Q_{t,i,j}^{flow}/S_{i,j}^{flow}$ | Active/Reactive/Apparent power flow between buses i and j |
| $P_{t,i}^{TES-Stir}/P_{t,i}^{in,TES}$ | Output/input power of TES |
| $Q_t^{Im,LEC}/Q_t^{Ex,LEC}$ | Imported/Exported reactive power to/from LEC |
| $Q_t^{TES}$ | Enthalpy of TES |
| $SOC_{t,i}$ | State of charge of BES |
| $T_t^{TES}$ | Temperature of phase change material (PCM) of TES |
| $T_{t,i}^{HW}$ | Hot water temperature of SB tank |
| $u_{t,i}^{ch,BES}/u_{t,i}^{d,BES}$ | Binary variables for charge/discharge of BES |
| $u_{t,i}^{ch,TES}/u_{t,i}^{dis,TES}$ | Binary variables for charge/discharge of TES |
| $u_{t,i}^{1}-u_{t,i}^{5}$ | Binary variables for linearizing TES characteristic |
| $W_{t,i}^{Stir}$ | Work output of the Stirling engine heat cycle |
| $W_{t,i}^{M,Stir}$ | Mechanical work output of the Stirling engine |
| $V_{t,i}$ | Voltage magnitude of bus |
| $z_{t,i}^{aux1}, z_{t,i}^{aux2}, z_{t,i}^{aux3}$ | Auxiliary variables indicating enthalpy of each region |

***Parameters***

| | |
|---|---|
| $\alpha_{t,i}^{cong}$ | Congestion parameter |
| $\epsilon^{H,Stir}$ | Effectiveness of the heat exchanger of Stirling engine |
| $\eta_d/\eta_{ch}$ | Charge/discharge efficiency of BES |
| $\eta^{h,TES}/\eta^{h,SB}$ | Efficiency of TES/SB heater |
| $\eta^{Gen}/\eta^{T}$ | Efficiency of TES output generator/ CHP turbine |
| $\eta^{W,Stir}$ | Efficiency of Stirling engine thermodynamic cycle |
| $\eta^{Mec,Stir}$ | Mechanical efficiency of Stirling engine |
| $\kappa_t^{DR}$ | Power to energy ratio of demand response (DR) |
| $\theta_{ij}$ | Impedance angle of branch between bus i and j |
| $\lambda_t^{e}$ | Electricity energy price |
| $\lambda_t^{h}$ | District heating variable heat price |
| $\xi^{DR}$ | Parameter corelating maximum energy capacity of DR as a percentage of load |
| $\xi^{l}$ | Coefficient of correlation of active and reactive power of CHP |
| $b_{ij}/Z_{ij}$ | Susceptance/ Impedance of branch between buses i and j |
| $c_1, c_2, c_3, c_4$ | Parameters of flexibility price |
| $c^{CHP}$ | Generation cost coefficient of boiler and CHP unit |
| $c^{deg}$ | Parameter to emulate degradation cost of BES |
| $C^{L}/C^{S}$ | Specific heat of PCM inside TES in liquid/solid phase |
| $C^{LS}$ | Load shedding cost |
| $C^{W}$ | Specific heat of water |
| $COP_i$ | Coefficient of performance of heat pump |
| $H^{Lat}$ | Latent heat of PCM in TES |
| $k/A$ | Thermal convection parameter/Area of SB tank |
| $M/M1$ | Big numbers in Big M method |
| $m^{TES}$ | Mass of PCM inside TES tank |
| $P_{i,t}^{l}/Q_{i,t}^{l}$ | Active/reactive demand |

The work of this paper was supported by funding from projects SUNSETS and GENTE within Solar ERANET and ERA-Net SES program, funded by the Swedish Energy Agency with project numbers 51197-1 and 52601-1. *Corresponding author: Maryam Mohiti.* M. Mohiti, M. Mazidi, D. Steen, T. Le are with Chalmers University of Technology, Gothenburg, Sweden (e-mail: [maryamar, mazadi, david.steen, tuan.le]@ chalmers.se).

| | |
|---|---|
| $P_{t,i}^{PV}$ | Forecasted PV power |
| $T_{t,i}^{a}$ | Ambient temperature |
| $T_{t,i}^{CW}$ | Temperature of cold water in SB tank |
| $T^{L1}/T^{L2},T^{S1}/T^{S2}$ | Temperature boundaries of PCM liquid and solid phase in TES |
| $T^{offset}$ | Offset value in enthalpy curve of PCM in TES |
| $T_{t}^{TES,0}$ | Initial Temperature of PCM inside TES |
| $RR^{B}$ | Ramp rate limit of CHP boiler |
| $V$ | Tank volume of SB |
| $V_{t,i}^{HW}$ | Hot water demand |
| Symbol | |
| $\overline{(\cdot)}/\underline{(\cdot)}$ | Maximum/minimum bounds of $(\cdot)$ |

# I. I. INTRODUCTION

Harnessing renewable energy sources (RESs) is vital for reducing carbon emissions and driving the transition to cleaner energy systems. The growing integration of distributed energy resources (DERs), such as flexible demands, energy storage systems (ESSs) and RESs is reshaping modern distribution networks. However, the unpredictability of RESs and flexible loads, along with bi-directional power flows, creates challenges such as grid congestion and voltage limits violations. Conventional centralized control methods for grid congestion management often require significant investment and long implementation timelines. DERs, however, can offer flexibility to support distribution system operators (DSOs) in ensuring reliable grid operation.

Congestion management in power systems can be broadly classified into two categories: technical and economic. While technical methods involve real-time physical interventions to resolve congestion, economic congestion management rely on market-based mechanisms to alleviate grid constraints [1]. Given that the technical methods are configured for real-time interventions, they are primarily employed as curative measures and the focus of this paper is on the latter. At the local level, leveraging flexibility has proven to be an effective tool for mitigating grid bottlenecks. According to [2], market-based congestion management mechanisms can be further divided into energy tariffs, capacity tariffs, and flexibility markets. Energy and capacity tariffs rely on price signals from DSOs to incentivize consumer behavior, whereas flexibility markets present a more dynamic and market-driven solution. By enabling DERs to actively participate and monetize their flexibility, these markets offer a framework for resolving grid congestion and enhancing grid resilience [3].

## A. State of art

The research conducted in [1-3] address congestion issues by controlling flexible sources through market-based mechanisms, either in a centralized or decentralized manner. However, in these studies, flexibility providers participate only in the LEM, and flexibility itself is not traded. Directive (EU) 2019/944 highlights the importance of DSOs offering market-based incentives for flexibility procurement [4]. The flexibility can be traded in long-term (capacity availability) and short-term (real-time or near real-time procurement) markets, known as LFCM and LFM, respectively. In [5], aggregators participate in both the LFM and LEM, receiving fixed remuneration from the DSO. Similarly, the authors of [6] propose a multi-DSO LFCM for trading energy and flexibility capacity. In [7], a platform for peer-to-peer energy and flexibility trading is introduced, designed to manage grid congestion in near real-time. Additionally, [8] explores the use of DERs for flexibility provision in both distribution and transmission networks. In these studies, the flexibility price is typically set by a DSO tariff, rather than being determined by the LFM. Other researches focus on flexibility pricing mechanisms, such as [9, 10]. In [9], the results of the EU Interflex project, which developed a market-based control mechanism for flexibility to relieve congestion on a transformer branch, are presented. The flexibility price is determined by the DSO which calculates it based on the cost of transformer lifetime reductions and the financial risk of an outage. The EU project CoordiNet [11] proposes a flexibility market that facilitates cooperation between DSOs and transmission system operators (TSO) with pricing mechanisms based on a pay-as-bid approach combined with nodal pricing. Similarly, the Erena project employs the pay-as-bid approach as the pricing scheme for the LFM [12]. [13] as part of the EU project FlexiGrid introduces a capacity limit product to be traded in the LFM and LFCM. The EU Bright project introduces a blockchain-based LFM with smart contracts between flexibility buyers and providers [14].

Most of the above works focus primarily on the bidding and clearing strategies of the flexibility market (FM) and lack incorporating network constraints or addressing how flexibility markets effectively mitigate congestion. Studies that do consider network constraints in their modeling of flexibility market are limited [9, 15, 16]. For example, the flexibility mechanism in [9] targets the relief of a transformer overloading but lacks scalability to all grid lines. In [16], a three-stage LFCM is introduced, where flexibility services are offered to both DSOs and TSOs. The LFCM operates in three stages: first, flexibility offers from prosumers are matched with the needs of DSOs and TSOs to maximize social welfare. Next, the DSO evaluates these matches against local network constraints for feasibility. Finally, the accepted offers are confirmed based on these checks. However, this study focuses solely on the LFCM and does not explore the interactions with the LEM. In the EU FlexGrid project, a continuous flexibility market is introduced where network constraints are integrated, and asymmetric block bids are employed for market clearing [17]. The market operates in near real-time but does not account for flexibility capacity. In [15], the authors propose a network-aware flexibility request, submitted to the LFM by the DSO. Similarly, [10] presents a network-aware clearing strategy for both LEM and LFCMs, where flexibility capacity is defined on a day-ahead basis, and the required flexibility is dispatched in real-time. In the aforementioned works, either the LEM are not modeled alongside the FM, or they are cleared sequentially rather than being jointly integrated.

To promote decentralized energy management, the European Commission has introduced the concept of LECs in EU legislation [18]. LECs are defined as legal entities that are based on voluntary participation and are effectively controlled by members aiming to provide environmental, economic, or social community benefits. LECs can actively help unlock the flexibility of DERs and solve congestion problems in the DN. LECs may consist of various energy carriers which the intrinsic coupling between these carriers enables more opportunities for participating in the FMs. Although many studies have focused on LECs offering flexibility in LEMs, [19-21] research on LEC participation in this area remains very limited. In [5], the optimal management of resources for a smart home aggregator, including TES, BES, and PVs, is addressed for participation in day-ahead energy market and LFCMs. However, this study is limited to a single aggregator, with the flexibility activation request provided as an input signal, and it does not account for the DN. The above work, in addition to their limitations in modeling flexibility market and its clearing strategies, does not consider DN constraints or congestion issues. Furthermore, it does not investigate how different heat sources, such as heat pumps and TESs in LECs, might affect their participation in FMs.

To the best of our knowledge, as summarized in Table I, there is a gap in literature concerning a framework that enables LECs to participate jointly in LEMs and FMs. This framework should take the network constraints into account to prevent congestion while allowing LECs to optimize their resources (i.e., DERs) for participation in energy, flexibility, and heat markets. Currently, there are no studies that adequately address this issue.

### B. *Paper Contributions and Structure*

In light of these knowledge gaps, this paper introduces a network-aware joint LEM and LFCM for LECs. The proposed preventive congestion management framework allows LECs to participate in the LFCM while utilizing the flexibility of their heat and electricity DERs to alleviate congestion in the DN. Within this framework, as the DSO receives the baseline dispatches from the LECs, it conducts a power flow analysis to identify congestion in the network and runs the LFCM on a day-ahead basis. LECs have the option to leverage their flexibility potential in the LEM, LFCM, or through heat exchange with the district heating system. In the proposed LFCM, LECs offer their flexibility capacity based on the bids received from the DSO and are compensated as the market clears. The main contributions of this paper are summarized as follows:

- Development of a preventive congestion management framework for LECs: The joint LEM and LFCM model enables LECs to either hold back their flexibility capacity and participate in the LFCM to relieve congestion or directly engage in the LEM and heat market.
- Network-aware and decentralized modeling: In response to growing concerns about privacy, the proposed congestion management framework is designed in a decentralized manner that incorporates network constraints.
- Dynamic flexibility price at LECs nodes: A dynamic pricing mechanism is proposed in the LFCM to determine the flexibility price traded between LECs and the DSO. Flexibility prices are updated in each iteration based on the location and severity of congestion, as well as the available flexibility from LECs. Once the process is complete, the LFCM clears the market by finalizing the flexibility bids.
- Preventing rebound effects: To avoid rebound effects—i.e., congestion occurring in another area or at a later time due to load shifts by LECs—the framework addresses this issue in the DSO's load flow analysis and constrains the capacity of LECs' power shifts.
- As part of the EU project SUNSETS [22], the performance of the proposed framework is demonstrated and evaluated using the testbed at Chalmers University of Technology in Sweden while some other DERs located in Greece are virtually connected through an IoT platform.

The remainder of the paper is organized as follows. In Section II the methodology of the proposed framework is discussed and in Section III the problem formulation is presented. The performance of the proposed framework is evaluated using the DN located at the Chalmers University of Technology campus in section IV, followed by the conclusion in Section V.

## II. METHODOLOGY

### A. *System model*

An overview of the preventive congestion management framework is illustrated in Fig.1. The proposed framework as part of the EU SUNSETS project is based on Sweden's energy market structure. In Sweden the energy market is deregulated, and the DSO is only responsible for the secure operation of the DN. The retailers which are independent stakeholders exchange energy (electricity and heat) with the LECs. In the proposed framework, the LECs can participate in the LEM, LFM, or exchange heat with the heat retailer. The LEM is cleared with the wholesale (spot) market price while the LFCM is activated by the DSO to alleviate congestion and flexibility price is cleared by the LFCM. The heat is exchanged with the heat retailer based on the district heating price which varies between months of the year. The capital flow between different stakeholders is also depicted Fig. 1 which will be later elaborated.

TABLE I
TAXONOMY TABLE OF LITERATURE

| Ref. | LEM | FM | | Heat market | Flexibility pricing | Network constraints | Flexibility providers | | | Congestion management | | |
|---|---|---|---|---|---|---|---|---|---|---|---|---|
| | | LFM | LFCM | | | | LECs | MG, aggregator | DERs | Preventive | Curative | Rebound |
| [6] | ✓ | ✓ | ✓ | – | Linear functions of the volume traded | ✓ | – | – | ✓ | ✓ | – | – |
| [7] | ✓ | ✓ | – | | Set from a range | ✓ | – | – | ✓ | ✓ | ✓ | – |
| [2] | ✓ | – | – | – | Transactive signals | – | – | ✓ | – | ✓ | – | – |
| [1] | ✓ | – | – | – | Fixed price | – | – | – | ✓ | – | – | – |
| [8] | ✓ | ✓ | – | – | Set from a range | ✓ | – | ✓ | – | – | – | – |
| [9] | ✓ | – | ✓ | – | Calculated based on transformer loss of life | – | – | ✓ | – | ✓ | – | – |
| [11] | – | ✓ | – | – | Market based and nodal pricing | ✓ | – | – | ✓ | – | – | – |
| [23] | – | ✓ | – | – | Smart contracts | – | – | – | ✓ | – | – | – |
| [16] | – | – | ✓ | – | Fixed price | ✓ | – | – | ✓ | – | – | – |
| [24] | ✓ | – | ✓ | – | Market based | ✓ | – | – | ✓ | – | – | ✓ |
| [15] | – | – | ✓ | – | Fixed price | ✓ | – | – | ✓ | – | ✓ | – |
| [10] | ✓ | ✓ | – | – | Nodal pricing | ✓ | ✓ | – | – | – | ✓ | – |
| [19, 21] | – | ✓ | – | – | Fixed price | – | ✓ | – | – | – | – | – |
| This paper | ✓ | – | ✓ | ✓ | Market based | ✓ | ✓ | – | – | ✓ | – | ✓ |

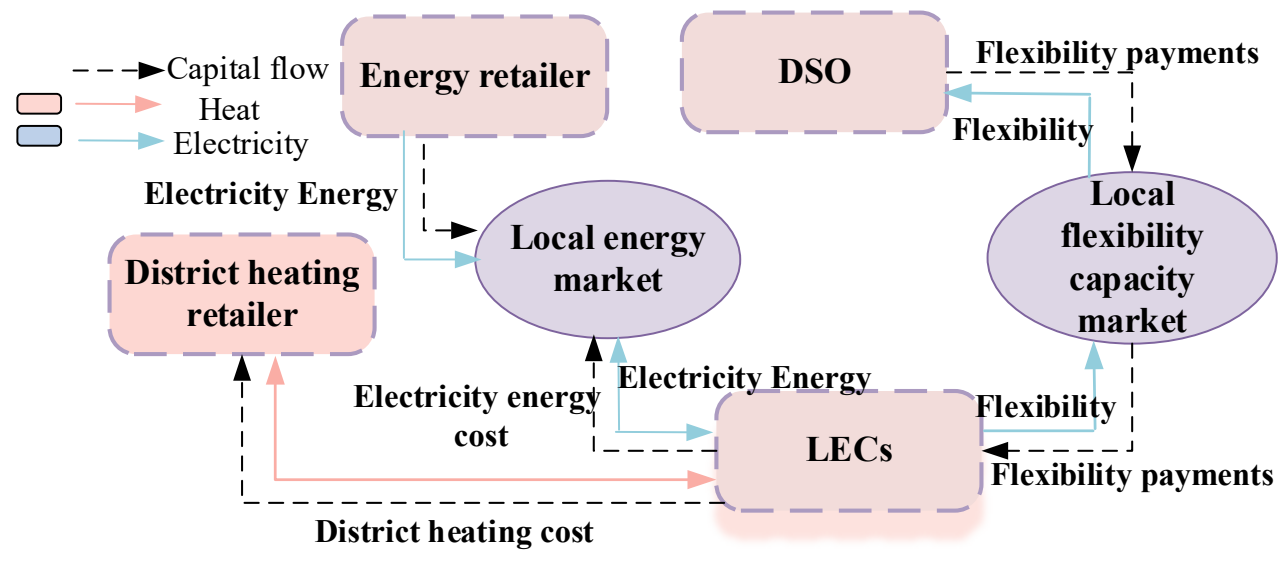

Fig. 1 Overview of the proposed framework

### B. *Local flexibility capacity market structure*

The LFCM is structured to trade flexibility capacities on a day-ahead basis and is activated by the DSO to prevent congestion in the DN. The working principle of the LFCM is as shown in Fig. 2. Initially, the LECs send their baseline schedules to the DSO. The baseline schedules of the LECs are optimized for minimizing their energy (electricity and heat) cost of load shifting, generation dispatch and energy price arbitrage in the day-ahead electricity market. Once the DSO receives the baseline schedules of the LECs, it conducts a power flow analysis to validate if the system is technically viable or if there is any congestion in the system. If there is congestion in the system the LFCM is activated, and the DSO determines the quantity of flexibility (active power at DN nodes) required in the network to alleviate the congestion.

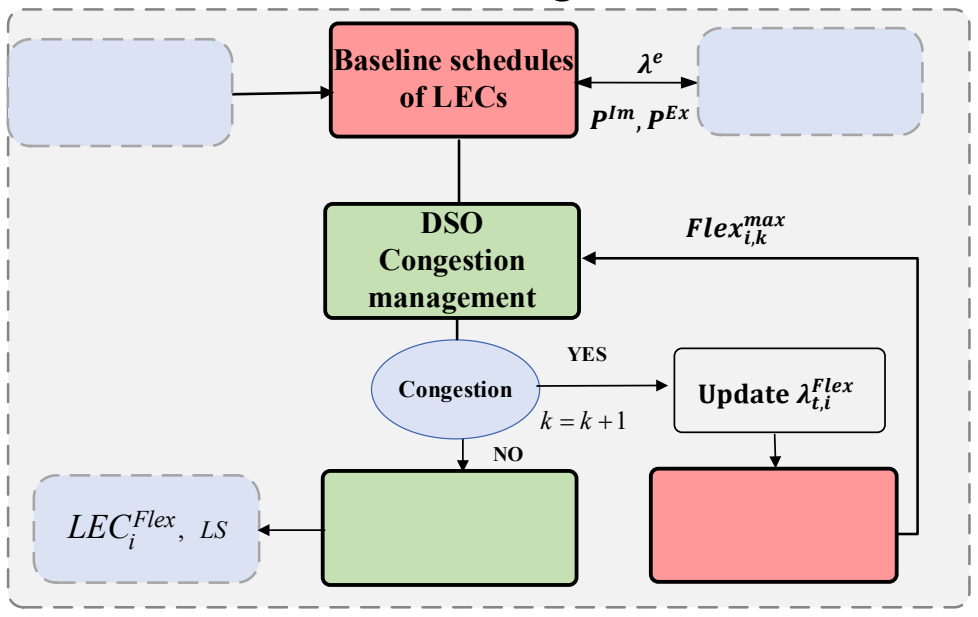

Fig. 2 Local flexibility capacity market structure

The DSO initializes the flexibility price ($\lambda_{t,i}^{Flex,k}$) by (1) and sends it to the LECs and based on this flexibility price, the LECs will bid their maximum flexibility. If the congestion is not relieved, the DSO will increase the price by the penalty factor $\rho_{t,i}^{Congestion,k}$ as presented in (2) in each iteration (k). According to (2), the penalty factor may decrease the congestion parameter ($\alpha_{t,i}^{cong}$) as the congestion gets relieved in each iteration. Note that the parameters of $c_1$-$c_4$ will only affect the initial value price and the speed of convergence of the algorithm, and these parameters can be manipulated based on the DSOs preferences. Accordingly, the flexibility price is updated based on the location and severity of congestion, as well as the available flexibility from LECs. As the algorithm proceeds, if the congestion is not relieved, the DSO increases the flexibility price and/or will perform load shedding. Once the congestion is relieved i.e., the amount of flexibility offered by the LECs and/or the load shedding is sufficient, the DSO runs the final flexibility allocation program and clears the market with the final accepted bids from each LEC.

$$\lambda_{t,i}^{Flex,k+1} = \lambda_{t,i}^{Flex,k} + \rho_{t,i}^{Congestion,k} \tag{1}$$

$$\rho_{t,i}^{Congestion,k} = c_1 \lambda_t^e (c_2 + c_3 e^{c_4 k})(\alpha_{t,i}^{cong}) \tag{2}$$

$$\alpha_{t,i}^{cong} = S_{i,j}^{max} - S_{i,j}^{flow} \tag{3}$$

## III. Mathematical Formulation of the Framework

In this section the mathematical formulation of the DSO and the LECs is presented.

### A. *DSO model*

The DSO is responsible for the secure operation of the DN. In the proposed framework the DSO runs the *congestion management* and the *Final congestion allocation* programs. Below these two problems are explained.

#### 1) *DSO congestion management problem*

In this problem, the DSO runs the power flow analysis and determines the amount of flexibility it requires to alleviate the congestion. The DN is connected to the upstream network at the substation network denoted as node 0.

*Objective function*: The DSO's objective is to minimize its costs of procuring flexibility and payment of load shedding as:

$$Min \sum_{t,i\in\mathbb{N}^{pcc}} \lambda_{t,i}^{Flex} FLEX_{t,i\in\mathbb{N}^{pcc}}^{max} + \sum_{t,i\in\mathbb{N}^{DN}} C^{LS} P_{i,t}^{LS} \tag{4}$$

*Power flow constraints:* The power flow equations of the DN are modelled as an AC power flow (5)-(9) and valid for the $\forall i \in \mathbb{N}^{DN}$. In (5) if the line is connected to the PCC of a LEC ($i \in \mathbb{N}^{pcc}$) the exchange power and the flexibility provided by the LEC are also incorporated in the power flow constraint. For the bus connected to the upstream network (i=0) the exchanged power with the upstream network is also considered in the power balance of the bus. Note that the line limits are not included in the power flow constraints of this problem and the congestion of the branches are checked with the parameter $\alpha_{t,i}^{cong}$ as in (3).

$$\sum_{j\in\mathbb{N}^{DN^i}} P_{t,i,j}^{flow} = P_{t,i}^{DG} + P_{t,i}^{PV} + P_{i,t}^{LS} + P_{i=0,t}^{grid} - P_{i,t}^{l} - P_{t,i\in\mathbb{N}^{pcc}}^{LEC} - FLEX_{t,i\in\mathbb{N}^{pcc}}^{max} , \tag{5}$$

$$\sum_{j\in\mathbb{N}^{DN^i}} Q_{t,i,j}^{flow} = Q_{t,i}^{DG} + Q_{i=0,t}^{grid} - Q_{i,t}^{l} - Q_{t,i\in\mathbb{N}^{pcc}}^{LEC} , \tag{6}$$

$$P_{t,i,j}^{flow} = \frac{V_{t,i}^2}{Z_{i,j}} \cos(\theta_{ij}) - \frac{V_{t,i}V_{t,j}}{Z_{i,j}} \cos(\delta_{t,i} - \delta_{t,j} + \theta_{ij}), \tag{7}$$

$$Q_{t,i,j}^{flow} = \frac{V_{t,i}^2}{Z_{i,j}} \sin(\theta_{ij}) - \frac{V_{t,i}V_{t,j}}{Z_{i,j}} \sin(\delta_{t,i} - \delta_{t,j} + \theta_{ij}) - \frac{b_{i,j}V_{t,i}^2}{2}, \tag{8}$$

$$\underline{V_{t,i}} \le V_{t,i} \le \overline{V_{t,i}} \tag{9}$$

$P_{t,i\in\mathbb{N}^{pcc}}^{LEC}$ and $Q_{t,i\in\mathbb{N}^{pcc}}^{LEC}$ represent the net baseline active and reactive power exchanged with LECs at their PCCs, as defined in (10). $FLEX_{t,i\in\mathbb{N}^{pcc}}^{max}$ denotes the maximum flexibility offered by the LECs. It is important to note that import/export power and flexibility are modelled as distinct variables, each subject to different pricing schemes. This separation allows for the joint modeling of the LFCM and LEM, and is further detailed in the *LECs flexibility scheduling* problem.

$$P_{t,i\in\mathbb{N}^{pcc}}^{LEC} = P_t^{Im,LEC} - P_t^{Ex,LEC} \quad (10)$$

$$Q_{t,i\in\mathbb{N}^{pcc}}^{LEC} = Q_t^{Im,LEC} - Q_t^{Ex,LEC} \quad (11)$$

*Flexibility price constraints:* The DSO updates the congestion penalty and flexibility price according to (1)-(2) and passes it to the LECs to determine their maximum amount of flexibility with the suggested price to relieve the congestion.

### 2) *DSO final flexibility allocation problem*

After the congestion is relieved in the *DSO congestion management problem*, this problem is solved to determine the amount of accepted flexibility from each LEC and the load shedding. Note this problem is the same as the *DSO congestion management problem* else that the amount of flexibility from each LEC is limited to the maximum offered flexibility from the LECs, and the branch limit constraints are considered.

*Objective function*: The objective of the DSO is the same as (4).

*Power flow constraints*: The power flow constraints are the same as the DSO congestion management problem, however, in this problem the branch limits i.e. (12) are considered to avoid congestion and the accepted amount of flexibility from each LEC ($LEC_{t,i\in\mathbb{N}^{pcc}}^{Flex}$) is determined by the DSO. Note that (12) is linearized with the method in [25].

$$-\overline{S_{t,i,j}^{flow}} \leq \sqrt{{P_{t,i,j}^{flow}}^2 + {Q_{t,i,j}^{flow}}^2} \leq \overline{S_{t,i,j}^{flow}} \quad (12)$$

*Flexibility constraints:* The accepted flexibility by the DSO should be lower than the maximum flexibility that the LECs have offered.

$$LEC_{t,i}^{Flex} \leq FLEX_{t,i}^{max}, i \in \mathbb{N}^{pcc} \quad (13)$$

## C. *Mathematical model of LECs*

The LECs optimize their operational problem in two stages of the framework. In the first stage they optimize their resources to submit their baseline schedules to the DSO which we refer to as the *LECs baseline scheduling problem*. If the DSO requires flexibility, in the congestion management block of Fig. 2, the LECs will receive the flexibility price from the DSO and they solve the LEC flexibility scheduling problem to submit their flexibility bids to the DSO. In the following, mathematical models of these problems are presented. In both problems, the LECs operators seek to optimally schedule their available DERs, while satisfying electricity and heat balance within the LEC and the operational constraints of the resources. The DERs that are available at Chalmers test site are CHP unit, PVs, BESS, DR resources, Azelio TES and the SB. The equations of this subsection are valid for the buses in the LECs, i.e., $\forall i \in \mathbb{N}^{LEC}$.

### 1) *LECs baseline scheduling problem*

*Objective Function*: In the baseline scheduling problem, the LECs tend to minimize their operation cost as presented in (14) to determine their baseline dispatches.

$$\begin{aligned} Min \sum_t \lambda_t^e\left(P_t^{Im,LEC} - P_t^{Ex,LEC}\right) + c^e P_t^{Im,LEC} \quad (14) \\ + \sum_t \lambda_t^h\left(H_t^{Im,LEC} - H_t^{Ex,LEC}\right) \\ + c^{H,var} H_t^{Im,LEC} + c^{H,fixed} \\ + \sum_{t,i} c^{CHP}\left(H_t^{B2DH} + H_t^T\right) \\ + \sum_{t,i} c^{deg}\left(P_{i,t}^{ch,BES} + P_{i,t}^{dis,BES}\right) \end{aligned}$$

The first term is the cost of exchanging active power with the DN. $\lambda_t^e$ is the energy price while $c^e$ is the network tariff (a fixed fee for utilizing the network) when the LEC imports energy. The second term is the cost of heat exchange with the district heating consisting of energy exchange cost. The district heating operator charges the district heating consumers by $c^{H,var}$ for the peak power input and $c^{H,var}$ which is a fixed annual cost (scaled to hourly value). The generation cost of the boiler and CHP unit is reflected by the third term. To avoid successive charge/discharge of the BESS the fourth term is added to emulate the degradation cost of the BESS. The TES has no degradation cost.

*Electrical energy balance:* The active and reactive power balances of the LECs are presented by:

$$\begin{aligned} P_{t,i}^{Ex,LEC} - P_{t,i}^{Im,LEC} = P_{t,i}^{CHP} + P_{t,i}^{PV} + P_{t,i}^{LS} + P_{t,i}^{DR} \quad (15) \\ + P_{t,i}^{dis,BES} + P_{t,i}^{TES-Stir} \\ - P_{t,i}^{ch,BES} - P_{t,i}^{in,TES} - P_{t,i}^{L} - P_{t,i}^{HP} \end{aligned}$$

$$Q_{t,i}^{Ex,LEC} - Q_{t,i}^{Im,LEC} = Q_{t,i}^{CHP} - Q_{t,i}^{l} \quad (16)$$

*Heat energy balance*: The heat energy balance of the LECs is presented as follows:

$$\begin{aligned} H_{t,i}^{Ex,LEC} - H_{t,i}^{Im,LEC} = H_{t,i}^{CHP} + H_{t,i}^{R,TES} + H_{t,i}^{HP} - H_{t,i}^{SB} \quad (17) \\ - H_{t,i}^{L} \end{aligned}$$

*Biomass boiler and CHP*: The biomass boiler and turbine can provide electricity and heating simultaneously to the electricity and district heating networks. The heat of the boiler can be dispatched directly to the district heating and to the turbine of the CHP unit presented as (18). The ramp rates of the boiler are shown with (19) while its output heat is limited by its operational limit (20). The output electricity of the turbine is reflected as (21) and coupled with the CHP output heat with (22). The active and reactive power of the CHP can be dispatched within their operational limits (23). The reactive capability of the turbine is coupled to its active power (24).

$$H_{t,i}^{B} = H_{t,i}^{B2DH} + H_{t,i}^{T} \quad (18)$$

$$-RR^B \leq H_{t,i}^{B} - H_{t-1,i}^{B} \leq RR^B \quad (19)$$

$$\underline{H_{t,i}^{B}} \leq H_{t,i}^{B} \leq \overline{H_{t,i}^{B}} \quad (20)$$

$$\overline{P_{t,i}^{CHP}} = \eta^T H_{t,i}^{T} \quad (21)$$

$$H_{t,i}^{CHP} = (1-\eta^T) P_{t,i}^{CHP} \quad (22)$$

$$\underline{P_{t,i}^{CHP}} \leq P_{t,i}^{CHP} \leq \overline{P_{t,i}^{CHP}} \quad (23)$$

$$\xi^l \overline{P_{t,i}^{CHP}} \leq Q_{t,i}^{CHP} \leq \xi^h P_{t,i}^{CHP} \quad (24)$$

*BES*: The operational constraints of the BES are considered as follow:

$$SOC_{t,i} = SOC_{t-1,i} - P_{t,i}^{d,BES}/\eta_d + P_{t,i}^{ch,BES}.\eta_{ch} \quad (25)$$

$$0 \leq P_{t,i}^{ch,BES} \leq u_{t,i}^{ch,BES}.\overline{P_{t,i}^{ch,BES}} \quad (26)$$

$$0 \leq P_{t,i}^{d,BES} \leq u_{t,i}^{d,BES}.\overline{P_{t,i}^{d,BES}} \quad (27)$$

$$u_{t,i}^{ch,BES} + u_{t,i}^{d,BES} \leq 1 \quad (28)$$

$$\underline{SOC_{t,i}} \leq SOC_{t,i} \leq \overline{SOC_{t,i}} \quad (29)$$

*Demand response:* The model of load flexibility is described by (30)-(32) and similar to an energy storage model [26]. The maximum energy capacity of the DR source is a percentage of the load at the LEC bus and the power to energy ratio parameter which depends on the technology of the DR resource.

$$E_{t,i}^{DR} = E_{t-1,i}^{DR} + P_{t,i}^{DR} \tag{30}$$

$$0 \leq E_{t,i}^{DR} \leq \xi^{DR} P_{t,i}^{L} \tag{31}$$

$$-\kappa_i^{DR}\xi^{DR}P_{t,i}^{L} \leq P_{t,i}^{DR} \leq \kappa_i^{DR}\xi^{DR}P_{t,i}^{L} \tag{32}$$

*TES:* Azelio TES is a latent phase thermal energy storage. The heat of the TES is transferred to a Stirling engine through a heat transfer fluid, on demand. The Stirling engine runs a generator for electricity and the low temperature (55-65℃ ) residual heat of the Stirling engine can be utilized through the heat to energy technology. The detailed model of the TES is presented in our previous work [27]. Here, we only provide the equations along with brief explanations. The enthalpy of the TES is presented as a function of the temperature as follows:

$$Q_t^{TES} = \int_{T_t^{TES,0}}^{T_t^{TES}} m^{TES}C^{TES}dT = m^{TES}C^{TES}\left(T_t^{TES} - T_t^{TES,0}\right) \tag{33}$$

Where depending on the temperature of the TES, the PCM can be in solid, liquid or latent phase and the specific heat, i.e., $C^{TES}$ is defined as follows:

$$C^{TES} = \begin{cases} C^{L} & T^{L1} < T_t^{TES} < T^{L2} \\ C^{Lat} & T^{S2} \leq T_t^{TES} \leq T^{L1} \\ C^{S} & T^{S1} < T_t^{TES} < T^{S2} \end{cases} \tag{34}$$

The solid and liquid specific heats, i.e., $C^S$ and $C^L$ are assumed constant while $C^{Lat}$ is derived from the slope of the enthalpy-temperature curve in [27] :

$$C^{Lat} = H^{Lat}/2T^{offset} \tag{35}$$

The enthalpy difference equals the charge and discharge heat, as well as idle losses, in the case of no charge/discharge. In discrete form, for the time step of Δt, this can be expressed as (36):

$$m^{TES}C^{TES}\left(T_{t+1,i}^{TES} - T_{t,i}^{TES}\right) = \left(H_{t,i}^{ch,TES} - H_{t,i}^{dis,TES} - H_{t,i}^{idle,TES}\right)\Delta t \tag{36}$$

The problem is converted to MILP with the big M method with the set of equations of (37)-(51) [27].

$$z_{t,i}^{aux1} + z_{t,i}^{aux2} + z_{t,i}^{aux3} = \left(H_{t,i}^{ch,TES} - H_{t,i}^{dis,TES} - H_{t,i}^{idle,TES}\right)\Delta t \tag{37}$$

$$-Mu_{t,i}^{1} \leq z_{t,i}^{aux1} \leq Mu_{t,i}^{4} \tag{38}$$

$$m^{TES}C^{L}\left(T_{t+1,i}^{TES} - T_{t,i}^{TES}\right) - M\left(1 - u_{t,i}^{4}\right) \leq z_{t,i}^{aux1} \leq m^{TES}C^{L}\left(T_{t+1,i}^{TES} - T_{t,i}^{TES}\right) + M\left(1 - u_{t,i}^{4}\right) \tag{39}$$

$$T^{L1} - M\left(1 - u_{t,i}^{4}\right) \leq T_{t,i}^{TES} < T^{L2} \tag{40}$$

$$-Mu_{t,i}^{5} \leq z_{t,i}^{aux2} \leq Mu_{t,i}^{5} \tag{41}$$

$$m^{TES}C^{Lat}\left(T_{t+1,i}^{TES} - T_{t,i}^{TES}\right) - M\left(1 - u_{t,i}^{5}\right) \leq z_{t,i}^{aux2} \leq m^{TES}C^{Lat}\left(T_{t+1,i}^{TES} - T_{t,i}^{TES}\right) + M\left(1 - u_{t,i}^{5}\right) \tag{42}$$

$$T^{S2} - M\left(1 - u_{t,i}^{5}\right) \leq T_{t,i}^{TES} \leq T^{L1} + M\left(1 - u_{t,i}^{5}\right) \tag{43}$$

$$-Mu_{t,i}^{1} \leq z_{t,i}^{aux3} \leq Mu_{t,i}^{1} \tag{44}$$

$$m^{TES}C^{Lat}\left(T_{t+1,i}^{TES} - T_{t,i}^{TES}\right) - M\left(1 - u_{t,i}^{1}\right) \leq z_{t,i}^{aux3} \leq m^{TES}C^{Lat}\left(T_{t+1,i}^{TES} - T_{t,i}^{TES}\right) + M\left(1 - u_{t,i}^{1}\right) \tag{45}$$

$$T^{S1} < T_{t,i}^{TES} \leq T^{S2} + M1\left(1 - u_{t,i}^{1}\right) \tag{46}$$

$$\left(T^{S2} - T_{t-1,i}^{TES}\right)/M1 \leq u_{t,i}^{1} \tag{47}$$

$$\left(T_{t-1,i}^{TES} - T^{S2}\right)/M1 \leq u_{t,i}^{2}$$

$$u_{t,i}^{1} + u_{t,i}^{2} = 1$$

$$T_{t,i}^{TES} - T^{L1}/M1 \leq u_{t,i}^{4} \tag{48}$$

$$u_{t,i}^{4} + u_{t,i}^{3} = 1 \tag{49}$$

$$u_{t,i}^{5} \leq u_{t,i}^{3}\ , u_{t,i}^{5} \leq u_{t,i}^{2} \tag{50}$$

$$u_{t,i}^{2} + u_{t,i}^{3} - 1 \leq u_{t,i}^{5} \tag{51}$$

The TES is charged through an electrical heater and its charge and discharge power operational limits are enforced by (52)-(56).

$$H_{t,i}^{ch,TES} = \eta^{h,TES}P_{t,i}^{in,TES} \tag{52}$$

$$P_{t,i}^{in,TES} \leq u_{t,i}^{ch,TES}\overline{P_{t,i}^{in,TES}} \tag{53}$$

$$0 \leq H_{t,i}^{ch,TES} \leq u_{t,i}^{ch,TES}\overline{H_{t,i}^{ch,TES}} \tag{54}$$

$$0 \leq H_{t,i}^{dis,TES} \leq u_{t,i}^{dis,TES}\overline{H_{t,i}^{dis,TES}} \tag{55}$$

$$u_{t,i}^{ch} + u_{t,i}^{dis} \leq 1 \tag{56}$$

The discharged heat of the TES is directed through a Stirling engine driving a DC generator which produces electricity. The residual heat of the Stirling Engine is transferred to the heat network through a heat exchanger (57)-(62).

$$W_{t,i}^{Stir} = \eta^{W,Stir}H_{t,i}^{dis,TES} \tag{57}$$

$$H_{t,i}^{R,Stir} = \epsilon^{H,Stir}H_{t,i}^{dis,TES} \tag{58}$$

$$W_{t,i}^{M,Stir} = \eta^{Mec,Stir}W_{t,i}^{Stir} \tag{59}$$

$$P_{t,i}^{TES-Stir} = \eta^{Gen}W_{t,i}^{M,Stir} \tag{60}$$

$$u_{t,i}^{dis,TES}\underline{H_{t,i}^{R,Stir}} \leq H_{t,i}^{R,Stir} \leq u_{t,i}^{dis,TES}\overline{H_{t,i}^{R,Stir}} \tag{61}$$

$$u_{t,i}^{dis,TES}\underline{W_{t,i}^{Stir}} \leq W_{t,i}^{Stir} \leq u_{t,i}^{dis,TES}\overline{W_{t,i}^{Stir}} \tag{62}$$

*Smart boiler:* The SB is a controllable electric boiler tank, with adjustable active power. An electrical heater (63) converts electricity into heat to supply the system. It is assumed that the water storage remains constantly full, with any hot water consumed being replaced by an equal volume of cold water during each time interval. The water storage temperature can be determined using equation (64) and is constrained by equation (65).

$$H_{t,i}^{SB} = \eta^{h,SB}.P_{t,i}^{SB} \tag{63}$$

$$T_{t+1,i}^{HW} = \left[V_{t,i}^{HW}.\left(T_t^{CW} - T_{t,i}^{HW} + V.T_{t,i}^{HW}\right]/V + \left[\left(H_{t,i}^{SB} \times 3600\right) - kA\left(T_{t,i}^{HW} - T_{t,i}^{a}\right)\right]/VC^{W} \tag{64}$$

$$\underline{T_{s,t}^{HW}} \leq T_{s,t}^{HW} \leq \overline{T_{s,t}^{HW}} \tag{65}$$

*Heat pumps*: The HPs located at Chalmers network are of air source types. The performance of the heat pumps is modelled using the coefficient of performance $COP_i$ as shown in (66), while the electrical energy consumption is constrained by the maximum capacity, as represented in (67).

$$H_{t,i}^{HP} = COP_iP_{t,i}^{HP} \tag{66}$$

$$P_{t,i}^{HP} \leq \overline{P_{t,i}^{HP}} \tag{67}$$

2) *LECs flexibility scheduling problem*

*Objective Function*: In LEC flexibility scheduling problem based on the flexibility price that the DSO offers, the LECs minimize their operation cost while offering flexibility to the DSO (68). The LECs reschedule their resources to participate in the LFCM, providing flexibility to the DSO, or to take part in the LEM and the heat market. $t^{CG}$ is the set of hours that congestion exists in the DN. The flexibility offered by the LEC is the change in the import/export power as compared to the baseline schedules as (69).

The objective function of this problem differs from that of (15) in the first two and the last term. The first two terms are the cost of rescheduled import/export electricity in non-congested and

$$Min \sum_{t \notin t^{CG}} \lambda_t^e \left(P_t^{Im,LEC} - P_t^{Ex,LEC}\right) + \sum_{t \in t^{CG}} \lambda_t^e \left(P_t^{Im,LEC} - P_t^{Ex_base,LEC}\right) + c^e P_t^{Im,LEC} + \sum_t \lambda_t^h \left(H_t^{Im,LEC} - H_t^{Ex,LEC}\right) + c^{H,var} H_t^{Im,LEC} + c^{H,fixed} + \sum_{t,i} c^{CHP} \left(H_t^{B2DH} + H_t^T\right) + \sum_{t,i} c^{deg} \left(P_{i,t}^{ch,BESS} + P_{i,t}^{dis,BESS}\right) - \sum_t \lambda_{t,i}^{Flex} FLEX_{t,i \in \mathbb{N}^{pcc}}^{max} \quad (68)$$

$$FLEX_{t,i \in \mathbb{N}^{pcc}}^{max} = P_t^{Ex,LEC} - P_t^{Ex_base,LEC} + P_t^{Im_base,LEC} - P_t^{Im,LEC} \quad (69)$$

congested hours while the last term is the gained revenue by offering flexibility. To make this clearer, consider that the LEC was importing 10 kW i.e., $P_t^{Im_base,LEC} = 10, P_t^{Ex_base,LEC} = 0$ in the baseline schedule of a congestion hour (the second term in (68)). After receiving the flexibility price from the DSO, it decreases its import power to $P_t^{Im,LEC} = 4$ kW and offers 6 kW of flexibility as in (69). Now with this schedule, 4 kW of the import power is the energy and traded in the LEM with energy price, while the 6 kW is traded in the LFCM with the flexibility price. In the objective function the first term is zero as it is not valid for the congested hours, the second term will be 4-0 and the amount of energy traded, while the last term will be 6 and the flexibility traded. The same is true when the LEC is exporting power in the baseline schedule. In many studies, e.g., [28] the LECs or microgrids reschedule their import/export power to relieve the congestion and the same price is set for the flexibility and the energy or the LFM and LEM are cleared separately and not co-optimized.

To avoid rebound effects — i.e., congestion occurring at a later time due to load shifts by LECs, the constraint (70) is enforced in the flexibility scheduling of the LEC. In (70), $\alpha_{t,i}^{cong}$ is the available capacity of branch $l$ at time $t$ which is calculated by the DSO as indicated in (3). The available capacity of the line is divided equally between the LECs.

$$\sqrt{\left(P_{t,k}^{Im,LEC} - P_{t,k-1}^{Im,LEC}\right)^2 + \left(Q_{t,k}^{Im,LEC} - Q_{t,k-1}^{Im,LEC}\right)^2} \le \alpha_{t,l}^{cong} \Big/ |\mathbb{N}^{LEC}| \quad (70)$$

The operational constraints of the DERs of the LEC are the same as the initial scheduling problem (15)-(67).

## IV. Demonstration Site and Case Studies

### A. *Description of demo site*

The electrical distribution network at Chalmers University is utilized to evaluate the proposed framework and is shown in Fig. 3. The campus features an 831 kW solar PV system, and the total electricity load demand varies between 2.5 MW and 6 MW [3]. The network operates at a voltage level of 10.5 kV/0.4 kV. Chalmers heat network is connected to the district heating system. Two areas within the network have been designated as LECs, as illustrated by the dotted lines in Fig. 3, capable of exchanging both electricity and heat with the upstream network. LEC1 currently includes resources such as PV, HP, BES, and a biofuel-powered CHP. LEC2's resources consist of PV, HP, BES, flexible loads, SB, and Azelio TES. Note that the TES and SB are connected virtually through the IoT platform and are located at Åmol, Sweden and Greece, respectively. The data for TES is retrieved from the actual model [27] and the rest of the DERs data is provided in Table II. The proposed framework is investigated in the test system of Fig. 3 to show its application in relieving congestion in the DN while providing the opportunity for the LECs to participate in the LEM and LFCM.

Two case studies are employed in which in case study I (Case I) the congestion is originated from overloading the network while in case study II (Case II) the congestion is caused by the reverse power flow of high PV generation. Most analysis is conducted on Case I for giving a better vision on the LFCM performance, while Case II investigates the capability of the framework to fulfil SUNSETS project's goal of increasing PV hosting capacity. The LECs can exchange electricity with the LEM with the Nord Pool spot market price. The network charge is retrieved from the local DSO [29] and is equal to 68 SEK/MWh. The simulations are carried out with the data of 14th April 2021 when both PV generation and heat loads are available. The heat price is collected from the heat retailer and for April it is 0.359 SEK/kWh and the tariffs $c^{H,var}$ and $c^{H,var}$ are 76.33 and 2.35 SEK/day, respectively [29]. PV generation and load consumption are predicted with the AI methods from our previous work [30].

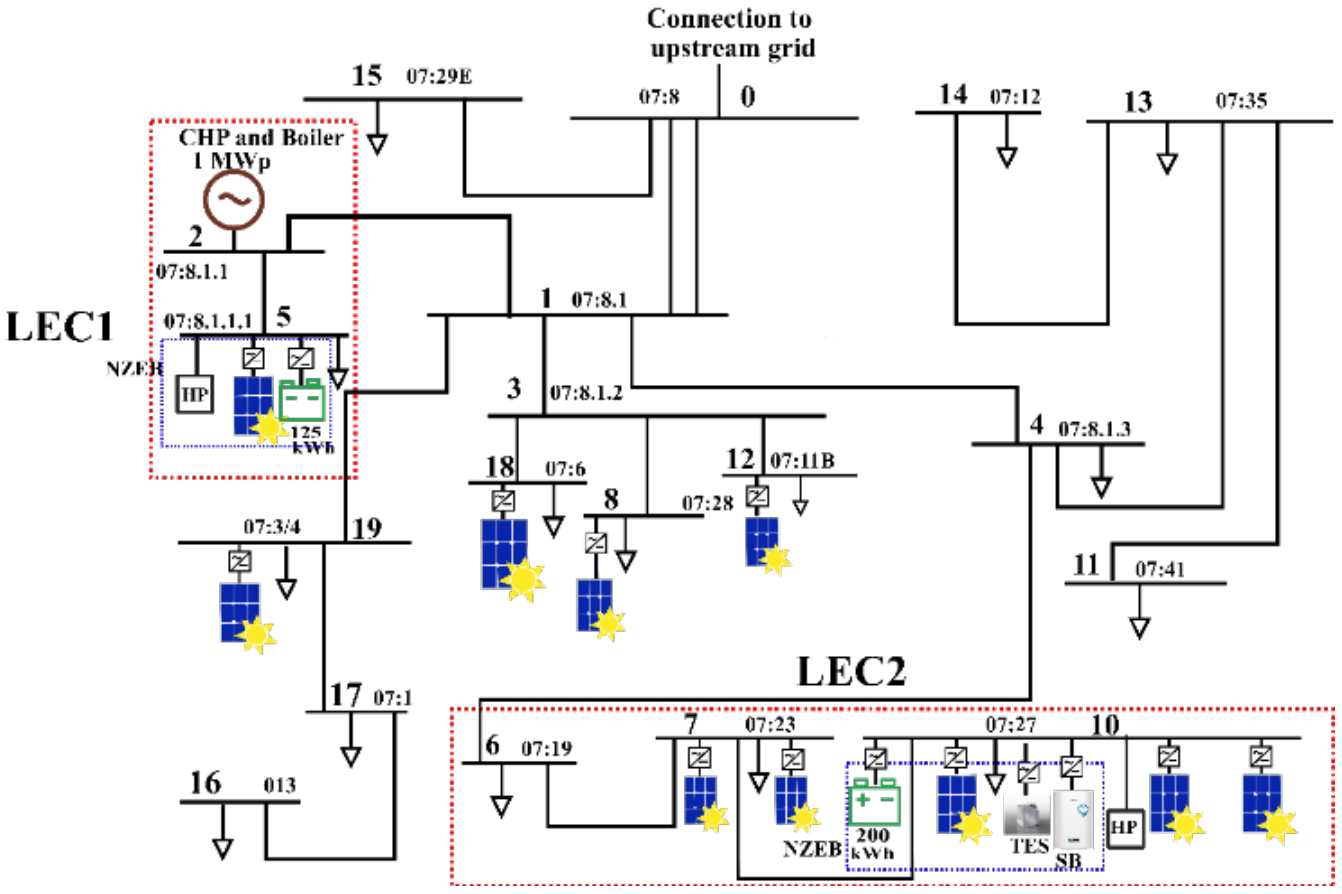

Fig. 3 Chalmers DN and the designated LECs

TABLE II
DATA OF THE DERS IN LECS

| Boiler and CHP unit | | | | BES |
|---|---|---|---|---|
| $\eta^B, \eta^T$ | $\xi^l, \xi^h$ | $RR^B, \underline{H_{t,i}^B}, \overline{H_{t,i}^B}$ | $\underline{P_{t,i}^{CHP}}, \overline{P_{t,i}^{CHP}}$ | $\eta^d, \eta^{ch}, c^{deg}$ |
| 0.77,0.17 | 0.2,0.8 | 100,200,1300 | 0,800 | 0.95,0.95, 0.01 |
| **BES** | | | **DR** | **HP** |
| $P_{t,i=10}^{ch,BES}, \overline{P_{t,i=10}^{d,BES}}$ | $\overline{P_{t,i=5}^{d,BES}}, \overline{P_{t,i=5}^{ch,BES}}$, | $\underline{SOC_{t,i}}, \overline{SOC_{t,i}}$ | $\xi^{DR}, \kappa_i^{DR}$ | $COP_{i=5}$, $COP_{i=10}$ |
| 100,100 | 50,50 | 0.2,0.9 | 0.2,0.5 | 3,4 |
| **HP** | **SB** | | | |
| $\overline{P_{t,i}^{HP}}$ | $kA, C_w, V, T_i^{CW}$ | $S_{t,i}^{SB}$ | $\underline{T_{s,t}^{HW}}, \overline{T_{s,t}^{HW}}$ | |
| 216 | 6, 4.18,600,10 | 10 | 60, 75 | |

### B. *Case I*

The performance of the LFCM to alleviate congestion is evaluated on Chalmers network. Under normal conditions, the network operates without congestion. However, to test the effectiveness of the LFCM, the rated capacity of the branch between bus 0 and bus 1 is artificially reduced by 82%, creating a congestion scenario. Once the framework is initialized, the LECs submit their baseline schedule (hourly electricity import/export) to participate in the day-ahead LEM. The DSO then performs a power flow analysis and identifies congestion on the branch between bus 0 and bus 1 during hours 10:00 to 16:00 and the LFCM is activated.

1) *Flexibility bids*

The LECs submit the maximum flexibility they are willing to offer with the suggested price by the DSO. Finally, the DSO runs the final flexibility allocation program and clears the LFCM market with the optimal amount of flexibility it should procure from each LEC.

The amount of flexibility required by the DSO, the offered flexibility bids from the LECs, the cleared flexibility by the LFCM, and the flexibility price are shown in Table III. To further clarify for instance; the total flexibility required from the DSO to relieve the congestion at hour 12:00 is 360.4 kW, which 235.9 kW and 124.5 kW are accepted from LEC1 and LEC2, respectively.

The maximum offered flexibility from LEC2 is 150.473 kW, however, this amount is not procured from it. Generally, it can be realized that in this case the DSO procures all the offered flexibility from LEC1 and in case the congestion is not relieved it procures the extra amount from LEC2. Since the congestion location (branch 1, i.e., branch between bus 0 and bus 1) is closer to the PCC of LEC1 (bus 2) and it will be more efficient to procure flexibility from LEC1.

The income of LEC1 increases by 40%, while the cost of LEC2 decreases by 3.45%, demonstrating financial benefits for both LECs.

TABLE III
FLEXIBILITY BIDS CASE I

| Hour | **Max. offered flexibility** (kW) | | **Cleared Flexibility by LFCM** (kW) | | **DSO flexibility requirement** (kW) | **Flexibility price** (SEK/kWh) |
|---|---|---|---|---|---|---|
| | LEC1 | LEC2 | LEC1 | LEC2 | | |
| 10:00 | 136.90 | - | 88.79 | - | 88.79 | 0.429 |
| 11:00 | 239.63 | 189.18 | 239.63 | 113.242 | 352.8 | 1.064 |
| 12:00 | 235.90 | 150.47 | 235.90 | 124.579 | 360.47 | 0.942 |
| 13:00 | 236.50 | 103.27 | 236.50 | 84.128 | 320.63 | 0.989 |
| 14:00 | 355.26 | 93.79 | 355.26 | 55.960 | 411.22 | 1.269 |
| 15:00 | 279.35 | 113.72 | 279.36 | 93.956 | 373.315 | 1.29 |
| 16:00 | 234.88 | 15.69 | 73.96 | - | 73.95 | 0.956 |

2) *Convergence*

The convergence of the flexibility price per iteration is depicted in Fig. 4. It can be traced, after 8 iterations the algorithm is converged and the congestion is relieved for all the congestion hours i.e., 10:00 to 16:00. However, for instance at hour 10:00 the market is cleared after 5 iterations. The congestion parameter ($\alpha_{t,i}^{Cong}$), flexibility price ($\lambda_{t,i}^{Flex,k}$) and the penalty factor ($\rho_{t,i}^{Cong,k}$) for hour 14:00 is illustrated in Fig. 5. The negative value of $\alpha_{t,i}^{Cong}$ indicates congestion in the branch. It can be followed from Fig. 5 congestion parameter is negative for the first 5 iterations and the congestion is relieved and $\alpha_{t,i}^{Cong}$ becomes non-negative in iteration 6. $\rho_{t,i}^{Cong,k}$ is the penalty factor which shows the amount of change in the flexibility price in each iteration.

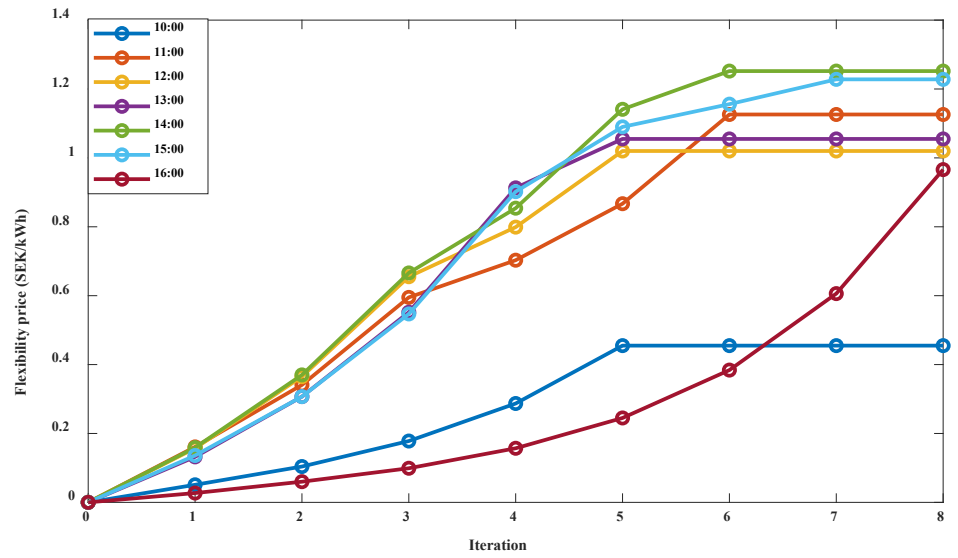

Fig. 4 Convergence of the flexibility price per iteration for each congestion hour

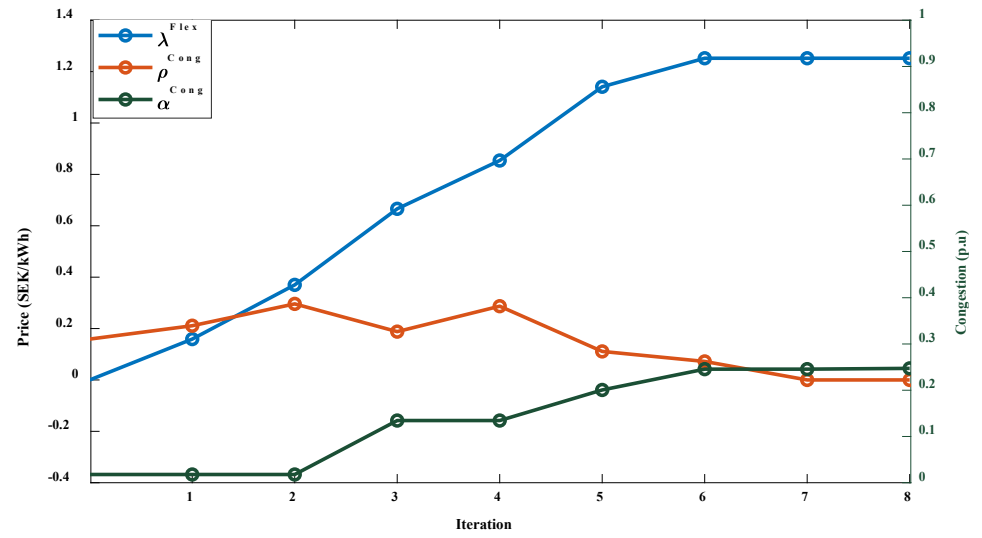

Fig. 5 Convergence of congestion parameter, flexibility price, and the penalty factor at hour 14:00

3) *Baseline and updated schedules of LECs participating in the LFCM*

The baseline and modified schedules of LEC1 and LEC2 after participating in the LFCM are shown in Fig. 6 and Fig. 7, respectively. The amount of flexibility provided by each DER and the total flexibility-i.e., the change in power exchange with the grid-is represented by dashed arrows. The direction of the arrows indicates whether the DER is reducing consumption or increasing generation to offer flexibility. LEC1 offers flexibility throughout the entire flexibility activation period, whereas LEC2 participates only during hours 11:00-13:00. At hour 15:00, DERs with shiftable dispatch capabilities, such as TES, BESS, and DR, tend to participate in the LFCM rather than the LEM due to flexibility's higher price compared to energy. For example, DR shifts its load curtailment from the morning peak price hours (9:00 and 10:00) to the congestion hours (12:00, 13:00, and 15:00). This highlights the importance of establishing a flexibility market, as it incentivizes LECs to adjust their baseline dispatch—originally optimized for minimizing costs in the energy market-in order to participate in the flexibility market.

## *C. Case II*

In this case, the framework's ability to alleviate congestion caused by the reverse power flow from RESs, specifically PV generation, is evaluated. Currently, the PV penetration in the Chalmers network is 5.6%, and no congestion is caused by reverse power flow from RESs. To simulate congestion, the network is modified by opening the line between bus 1 and bus 3 and connecting bus 3 to bus 4, which is feasible due to an available feeder at bus 4. PV penetration is increased from 5.6% to 65% by adding PV capacity at buses 8, 12, and 18. To further emulate a congestion event, the capacity of the branch between bus 1 and bus 4 is reduced by 60%.

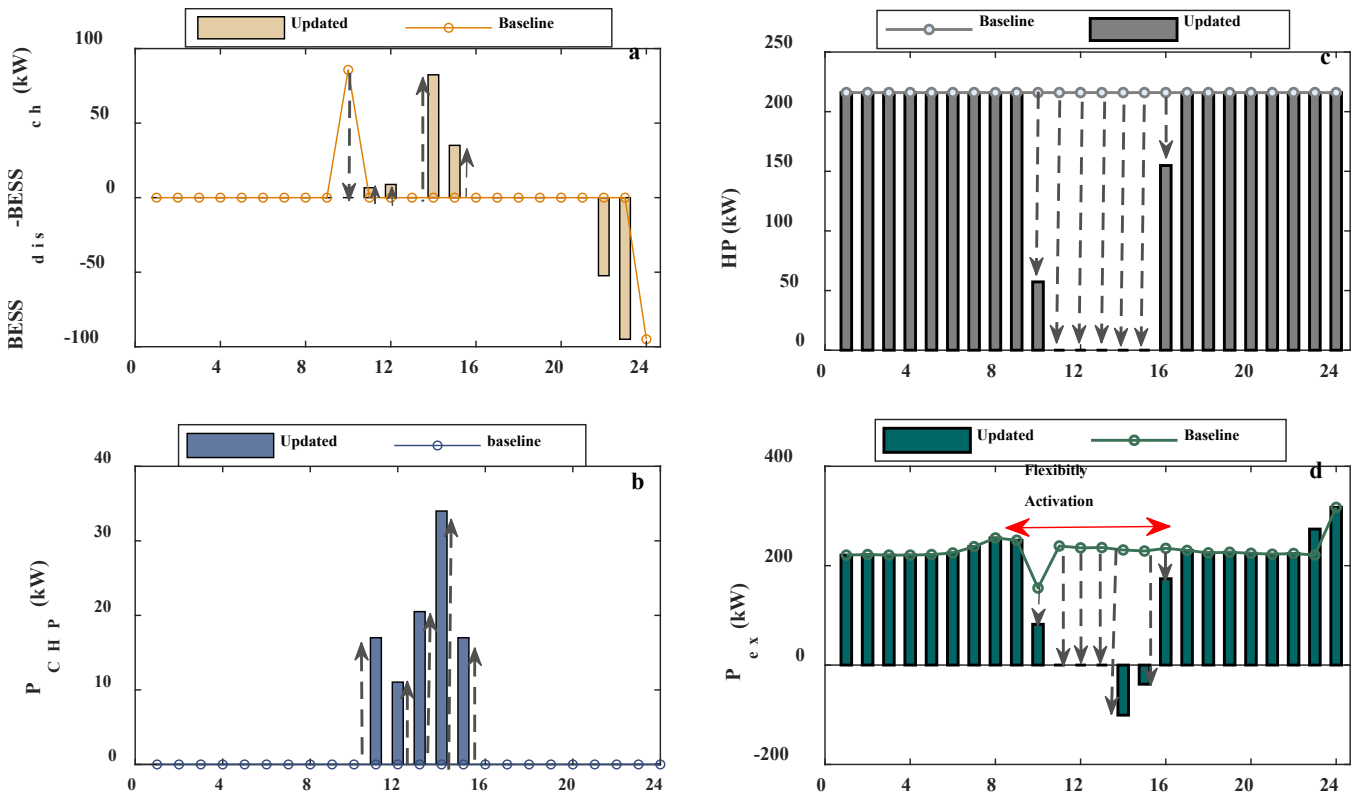

Fig. 6 Offered flexibility by each DER of LEC1(a-c) and total flexibility (d) in case study 1

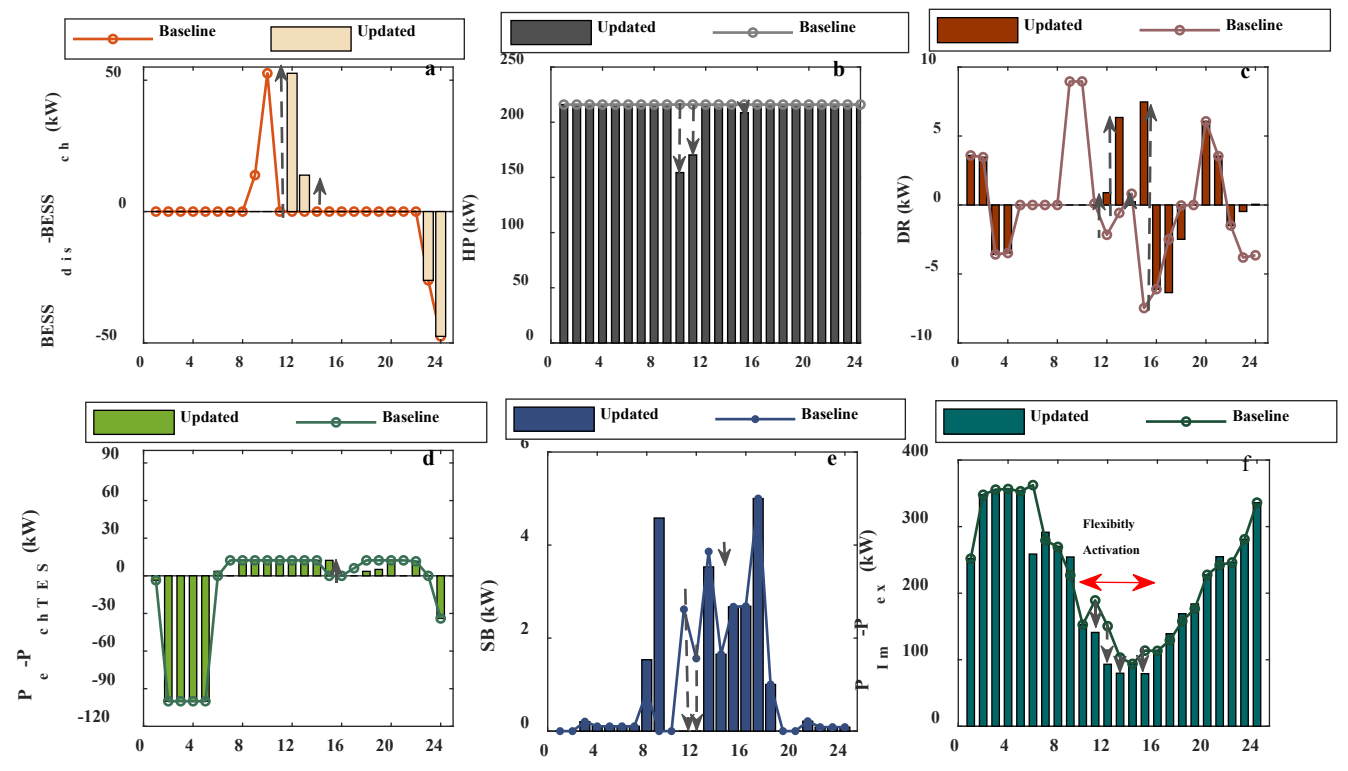

Fig. 7 Offered flexibility by each DER of LEC2(a-e) and total flexibility with grid (f) in case study I

### 1) *Flexibility bids*

At hour 14:00, after the framework is activated and LECs submit their baseline schedules, this line becomes congested due to PV generation in the lower feeders. The LFCM is activated, and the DSO requests flexibility from the LECs. Following the process shown in Fig. 2, the LECs submit their flexibility bids, and after the price is settled, the DSO allocates flexibility as outlined in Table 3. Since the congestion occurs on the branch between bus 1 and bus 4, flexibility from LEC1 cannot be used to relieve it, so the DSO only accepts flexibility bids from LEC2. By implementing the LFCM, PV penetration in the network increases by 59.4% without PV curtailment, meeting one of the key objectives of the EU SUNSETs project. LEC1 offers 95 kW of flexibility at hour 14:00 which is originated by changing the baseline schedule of its BES and charging in this hour. However, the flexibility offered is not accepted and not depicted.

### 2) *Baseline and updated schedules of LECs*

In Fig. 8 the baseline and updated schedules of the DERs of LEC2 is depicted. It can be seen the BES, TES charge in this hour and the DR shifts its curtailment from hour 14:00 to outside the flexibility activation period. The smart boiler also increases the hot water temperature to consume more electricity. It can be noticed that the HP did not participate in the flexibility provision as in the baseline schedule it already is consuming its maximum demand. Note that LEC1 flexibility bid is not accepted and hence the baseline and updated schedules of it are not depicted in Fig. 8.

TABLE III

FLEXIBILITY BIDS IN CASE 2

| Hour | Max. offered flexibility (kW) | | Cleared flexibility (kW) | | DSO flexibility requirement (kW) | Price of flexibility (SEK/kW) |
|---|---|---|---|---|---|---|
| | LEC1 | LEC2 | LEC1 | LEC2 | | |
| 14:00 | 95 | 172.141 | - | 161.992 | 161.992 | 0.229 |

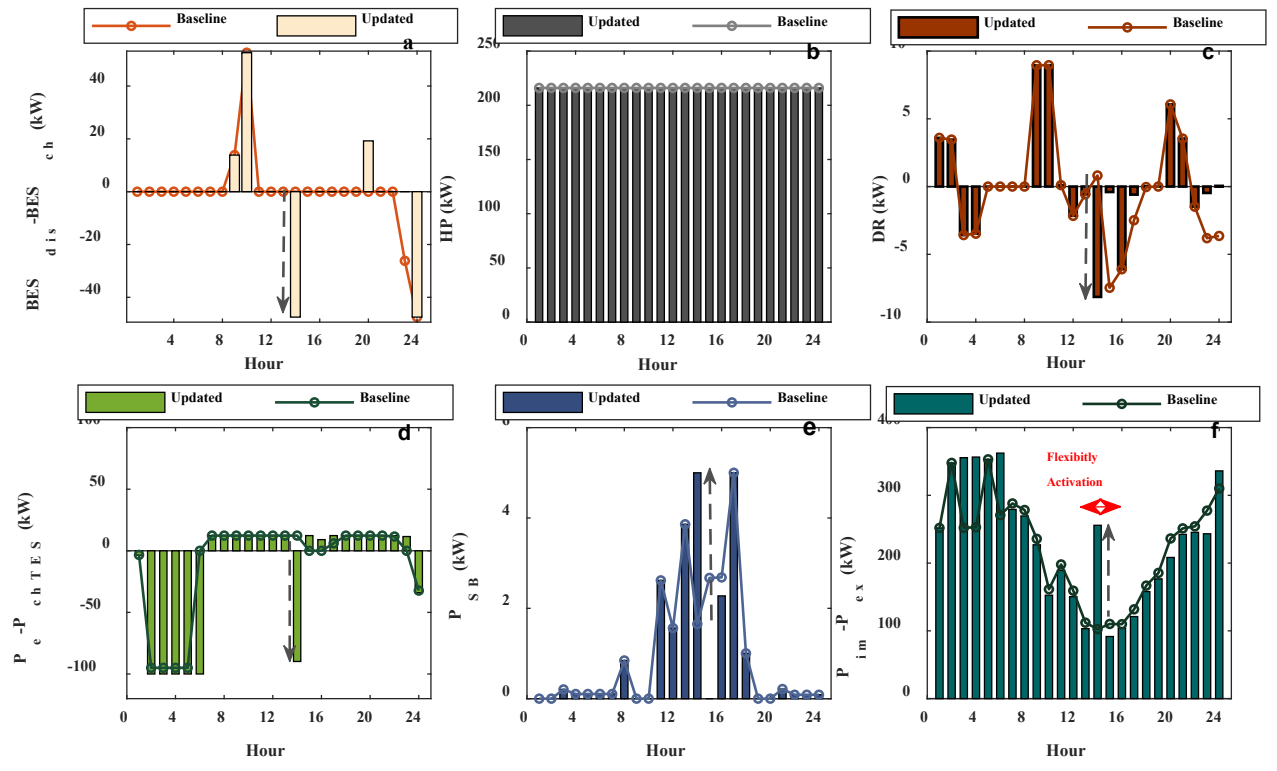

Fig. 8 Offered flexibility by each DER of LEC2(a-e) and exchanged power with grid (f) in case II

### 4) *Financial settlement*

In this section the financial settlement of the proposed framework is carried out. In Fig. 2, the capital flow of the financial statement is illustrated. As can be seen by participating in the LEM the energy cost is paid to the energy retailers while the DSO pays for flexibility. It should be noted that if the market is fully unbundled then the DSO cannot have direct capital flow with the LECs and a local aggregator/retailer may be required for the flexibility service. In Table IV, the cost and income changes of different stakeholders, with respect to the baseline is presented for case I. The operational cost change of LECs refers to the difference of all terms in the objective function except the flexibility income term in the *initial and flexibility scheduling problems*. It can be seen that the LECs gain revenue by participating in the flexibility market. The energy retailer's income decreases since the LECs shift part of their capacity from the LEM to the LFCM while the district heating retailer's income increases since the LECs turn off/ reduce their heat pumps electricity consumption to participate in the LFCM and they need to procure the decreased heat generation of the heat pumps from the district heat retailer. The financial capital that is originated from the increase in the DSO costs (1752 SEK) consists in the financial flows that goes to the energy retailer (-710.22 SEK), heat retailer (1425.218 SEK), the increment of the operational cost of the LECs (259.77 SEK), and the total revenue they earn by participating in the LFCM (777.51 SEK).

TABLE IV

COST AND INCOME CHANGE OF STAKEHOLDERS WITH RESPECT TO BASELINE FOR CASE 1

| Capital flow (SEK) | DSO | LEC | | Energy retailer | District heating retailer |
|---|---|---|---|---|---|
| | | LEC1 | LEC2 | | |
| **Operational cost increase** | 1752.26 | 259.77 | 0 | - | - |
| **Income increase** | - | - | - | -710.22 | 1425.218 |
| **Revenue** | - | 720.35 | 57.16 | - | - |

## V. Conclusion

This paper presents a novel network-aware LEM and LFCM framework for LECs. The framework enables LECs to leverage the flexibility of their heat and electricity DERs to participate in both the LEM and the LFCM. The proposed framework was tested in a real-world case study at Chalmers University of Technology in Sweden, with actual and virtual DERs connected via an IoT platform. In Case 1, where congestion was caused by network loading, the cleared flexibility prices were higher than energy prices, incentivizing LECs to shift resources from the heat market and LEM to the LFCM. In Case 2, where congestion was caused by reverse power flow, flexibility prices were lower due to the adequate storage capacity available to absorb surplus PV generation and participate in the LFCM. The findings show that the framework effectively alleviates congestion while providing financial benefits to LECs. The framework increased PV penetration by 59.4% without curtailment, meeting a key goal of the EU SUNSETs project. Financial analysis revealed that capital flows directly from energy and heat retailers to LECs participating in the LFCM providing up to 40% financial benefit for them.